# Doubly Cognitive Architecture Based Cognitive Wireless Sensor Network

Sumit Kumar, Deepti Singhal, Rama Murthy Garimella,
International Institute of Information Technology, Hyderabad, India - 500 032,
sumit.kumar@research.iiit.ac.in, deepti.singhal@research.iiit.ac.in, rammurthy@iiit.ac.in

*Abstract*—Nowadays scarcity of spectrum availability is increasing highly. Adding cognition to the existing Wireless Sensor Network (WSN) infrastructure will help in this situation. As sensor nodes in WSN are limited with some constrains like power, efforts are required to increase the lifetime and other performance measures of the network. In this paper we propose the idea of Doubly Cognitive WSN. The basic idea is to progressively allocate the sensing resources only to the most promising areas of the spectrum. This work is based on Artificial Neural Network as well as on Support Vector Machine (SVM) concept. As the load of sensing resource is reduced significantly, this approach will save the energy of the nodes, and also reduce the sensing time dramatically. The proposed work can be enhanced by doing the pattern analysis thing after a sufficiently long time again and again to review the strategy of sensing. Thus Doubly Cognitive WSN will enable current WSN to overcome the spectrum scarcity as well as save the energy of the sensor nodes.

## I. INTRODUCTION

Cognitive networking speaks for an intelligent communication system, consisting of both the wire line and/or the wireless connections, that is aware of its transmission environment, both internal and external, and acts adaptively and autonomously to attain its intended goal(s). This implies that all the network nodes and the end devices are self-aware and context-aware all of the time. The interest in cognitive networking is mainly driven from the need to manage the increasing complexity and the efficient utilization of available resources to deliver applications and services as economically as possible [1], [2].

A WSN is one of the areas where there is highest demand for cognitive networking. There are several reasons among which the recourse constraints (spectrum and power) are the most appealing one. Although in WSN the nodes are constrained in resources mainly in terms of battery power but these days there is scarcity increasing in terms of spectrum availability also. Traditionally the WSN work in ISM band (2.4 GHz), but in the same band we have many competing technologies working simultaneously like WLAN 802.11 a/b/g and ZigBee (802.15.4), Wi-Fi, Bluetooth. Hence in an environment where all these competing technologies are working simultaneously, it becomes difficult to find free spectrum to transmit without an error. Also at the same time the licensed mobile communication bands are almost free for 85% of the time w.r.t. spatial and temporal terms [3], [4].

Hence there can be two motives: either to find a free channel in the unlicensed band and do wireless transmission or to find a free channel in the licensed band and do communication. Also it will be strictly needed that whenever the licensed user comes back the cognitive user backs-off from the channel and switch to another free channel without creating any difficulty to the primary user. One another important thing is that if the cognitive user finds several free channels then it can go for the best available channel to do communication.

Another important scheme is there in which both the primary and secondary exist simultaneously as long as the QoS of the primary user is not compromised. If the interference created from the power transmitted by the secondary user still remains below some threshold, then this scheme can be helpful.

Adding cognition to the existing WSN infrastructure will bring about many benefits. CWSN will enable current WSN to overcome the spectrum scarcity as well as node energy problem. The cognitive technology will not only provide access to new spectrum but with better propagation characteristics too. Also by adaptively changing the systems parameters like modulation schemes, Transmit power, carrier frequency, constellation size a wide variety of data rates can be achieved. This will certainly improve the power consumption and network life time in a WSN. It will also help in coping with the fading (frequency selective/flat).

There are two important terms in this regard, **Cognitive Radio based WSN** and **Cognitive WSN**, which appear similar but are quite different from each other. A Cognitive Radio WSN is a WSN with each node having cognitive radio capability and nothing more than that. This means it will have cognitive capabilities in the physical layers only. But this will not fulfil our purpose as the WSN demands cognitive radio capabilities with networking among them which can take benefit from this cognition. Because contemporary protocols for layers above the physical layer are written for a fixed channel/bandwidth/QoS assignment schemes. Changing the physical layer parameter frequently will also affect the upper layer function. Hence for a scenario where the channel frequency as well as the allotted bandwidth will be changing too frequently a more flexible protocol will be required. Hence the concept for Cognitive WSN which involves cognition not only in the physical layer but will be a cross layered approach.

## II. PROPOSED APPROACH

Our work is based on pattern recognition as well as "Two-stage Spectrum Sensing for Cognitive Radios". Underlying notion of our idea is to progressively allocate the sensing resources to only the most promising areas of the spectrum. This translates in a reduction of sensing resources and time needed to accurately identify spectrum holes, in contrast with more conventional approaches that allocate the sensing budget over the entire spectrum uniformly. And this is needed in a WSN where there is a very hard constraint for energy as well as spectrum. Doing sensing only in the most promising area will not only save the energy of the spectrum sensing nodes (cluster heads) but also save the time to sense and react and transmit/broadcast the information to other nodes. Further enhancement in the technology in this area may also enable WSN to do sensing at node level instead of getting the information through cluster heads.. This idea will work very well in the scenarios where there is a multichannel cognitive radio network (it generally happens in WSN working along with WLAN, Wi-Fi, and Bluetooth). Hence multiple spectrum bands are sensed to identify transmission opportunities.

In this paper we propose the idea of **Doubly Cognitive WSN**. It is different in the sense that we propose a work which will be based on Artificial Neural Network as well as on SVM (support vector machine) concept. SVM is a kind of supervised learning methods that analyze data, recognize patterns, and used for pattern classification. The idea is very simple in the sense that it will begin by continuously collecting the spectrum white space pattern, traffic pattern, fading pattern, SNR pattern on each channel spatially/temporally for a sufficiently long time in the area in which the CWSN is proposed to be deployed. Once we will have a long time repository of the above data then it can be used as a training pattern for supervised learning of the **cognitive spectrum sensing machine** which can be trained to sense for the spectrum only in the bands or channels where there is a pattern of less traffic spatially/temporally and/or there is a pattern of having white space spatially/temporally. This approach will not only save the energy recourses of the nodes but also will reduce the sensing time dramatically. The cognitive spectrum sensing machine can become more effective by training it with respect to the fading and SNR pattern of the available channels. Then accordingly the radio parameters can be optimised w.r.t modulation scheme, constellation size, transmit power. Hence in brief the idea of doubly cognitive speaks about first learning from the pattern of the ambient w.r.t various transmission parameters (which is the first step of cognition) and then cognitively (intuitively) setting the system parameters to achieve the optimum performance [5].

The proposed work can be enhanced by doing the pattern analysis thing after a sufficiently long time again and again to review the strategy of sensing.

## III. ARCHITECTURAL INNOVATIONS FOR ENERGY AWARENESS

Instead of doing wide band spectrum sensing at the base station which consumes too much energy the task of sensing can be divided in between the cluster-heads so that each of them will be sensing a chunk of the wideband spectrum. These nodes can then share (co-operate) the spectrum sensing information with each other to create whole white space information. This sensing is done on the bands which are most likely to be vacant or having less traffic as predicted by the neural network algorithms. This will further reduce the sensing time and energy.

This division of task of spectrum sensing can be allotted to the cluster heads and they will share this information with each other and finally all of them will be having the information of the whole white space. Whenever there is query from any of the nodes within a cluster, the cluster head will see from the latest updated white space information and will direct the node to do communication over that free channel. While the communication is taking place, the cluster heads can keep themselves busy in updating their white space information. If in between the communication there is some change in the white space information over the channel which is engaged by the node doing communication then:

- The cluster head can direct that node to back-off form that channel and can direct him to a new channel as per availability.
- If no channels are available then the data has to be kept stored in the node until a channel becomes free to communicate.

## IV. CHALLENGING RESEARCH PROBLEMS

There are many challenges in developing a cognitive radio for WSN. The cognitive radio for a WSN has to be different in many aspects from a cognitive radio for mobile communication. First there will be a very tight constraint of battery life. As the computing capability of the nodes is far less than that of the cognitive radio devices used for mobile communication. Hence to implement reconfigurable nodes will be big deal apart from implementing the sensing algorithms. Even implementing neural network algorithms in cluster heads will be a big deal.